\documentclass {article}
\usepackage{amsmath}
\usepackage[pdftex]{graphicx}

\title{Coevolutionary Genetic Algorithms for Establishing Nash Equilibrium in Symmetric Cournot Games}
\author{Mattheos Protopapas\footnote{Department of Statistics, University of Rome ``La Sapienza'', Aldo Moro Square 5, 00185 Rome Italy. tel. +393391457307, e-mail: matteo.protopapas@uniroma1.it} \and Francesco Battaglia\footnote{Department of Statistics, University of Rome ``La Sapienza'', Aldo Moro Square 5, 00185 Rome Italy. tel. +390649910440, e-mail: francesco.battaglia@uniroma1.it}\and Elias Kosmatopoulos\footnote{Department of Production Engineering and Management, Technical University of Crete, Agiou Titou Square. tel. +302821037306. e-mail: kosmatop@dssl.tuc.gr}}

\begin{document}

\maketitle
\textbf{Abstract}. We use co-evolutionary genetic algorithms to model the players' learning process in several Cournot models, and evaluate them in terms of their convergence to the Nash Equilibrium. The ``social-learning'' versions of the two co-evolutionary algorithms we introduce, establish Nash Equilibrium in those models, in contrast to the ``individual learning'' versions which, as we see here, do not imply the convergence of the players' strategies to the Nash outcome. When players use ``canonical co-evolutionary genetic algorithms'' as learning algorithms, the process of the game  is an ergodic Markov Chain, and therefore we analyze simulation results using both the relevant methodology and more general statistical tests, to find that in the ``social'' case, states leading to NE play are highly frequent at the stationary distribution of the chain, in contrast to the ``individual learning'' case, when NE is not reached at all in our simulations; to find that the expected Hamming distance of the states at the limiting distribution from the ``NE state'' is significantly smaller in the ``social'' than in the ``individual learning case''; to estimate the expected time that the ``social'' algorithms need to get to the ``NE state'' and verify their robustness and finally to show that a large fraction of the games played are indeed at the Nash Equilibrium.

\vspace{0.3 cm}
\textbf{Keywords}: Genetic Algorithms, Cournot oligopoly, Evolutionary Game Theory, Nash Equilibrium.

\newpage

\section {Introduction}
The ``Cournot Game'' models an oligopoly of two or more firms that simultaneously define the quantities they supply to the market, which in turn define both the market price and the equilibrium quantity in the market. Co-evolutionary Genetic Algorithms have been used for studying Cournot games, since Arifovic [3] studied the cobweb model. In contrast to the classical genetic algorithms used for optimization, the co-evolutionary versions are distinct at the issue of the objective function. In a classical genetic algorithm the objective function for optimization is given before hand, while in the co-evolutionary case, the objective function changes during the course of play as it is based on the choices of the players. So the players' strategies and, consequently, the genetic algorithms that are used to determine the players' choices, co-evolve with the goals of these algorithms, within the dynamic process of the system under consideration. Arifovic (1994) used four different co-evolutionary genetic algorithms to model players' learning and decision making: two single-population algorithms, where each player's choice is represented by a single chromosome in the population of the single genetic algorithm that is used to determine the evolution of the system, and two multi-population algorithms, where each player has its own population of chromosomes and its own Genetic Algorithm to determine his strategy. Arifovic links the chromosomes' fitness to the profit established after a round of play, during  which the algorithms define the active quantities that players choose to produce and sell at the market. The  quantities chosen define, in turn, the total quantity and the price at the market, leading to a specific profit for each player. Thus, the fitness function is dependent on the actions of the players on the previous round, and the co-evolutionary "nature" of the algorithms is established. 

In Arifovic's algorithms [3], as well as any other algorithms we use here, each chromosome's fitness is proportional to its profit, as given by
\begin{equation}\label{profitEq}
\pi (q_i) = Pq_i-c_i(q_i)
\end{equation}
where $c_i(q_i)$ is the player's cost for producing $q_i$ items of product and P is the market price, as determined by all players' quantity choices, from the inverse demand function
\begin{equation}
P=a-b\sum_{i=1}^n q_i
\end{equation}
In Arifovic's algorithms, populations are updated after every single Cournot game is played, and converge to the Walrasian (competitive) equilibrium and not the Nash equilibrium [2],[14]. Convergence to the competitive equilibrium means that agents' actions -as determined by the algorithm- tend to maximize \eqref{profitEq}, with price regarded as given, instead of 
\begin{equation}\label{nash}
\max_{q_i}\pi(q_i)=P(q_i)q_i-c_i(q_i)
\end{equation}
that gives the Nash Equilibrium in pure strategies [2]. Later variants of Arifovic's model [5],[7] share the same properties.

Vriend was the first to present a co-evolutionary genetic algorithm in which the equilibrium price and quantity on the market -but not the strategies of the individual players as we will see later- converge to the respective values of the Nash Equilibrium [15]. In his individual learning, multi-population algorithm, which is one of the two algorithms that we study -and transform- in this article, chromosomes' fitness is calculated only after the chromosomes are used in a game, and the population is updated after a given number of games are played with the chromosomes of the current populations. Each player has its own population of chromosomes, from which he picks at random one chromosome to determine its quantity choice at the current round. The fitness of the chromosome, based on the profit acquired from the current game is then calculated, and after a given number of rounds, the population is updated by the usual genetic algorithm operators (crossover and mutation). Since the populations are updated separately, the algorithm is regarded as individual learning. These settings yield Nash Equilibrium values for the total quantity on the market and, consequently, for the price as well, as proven by Vallee and Yildizoglou [14].

Finally Alkemade et al. [1] present the first (single population) social  learning algorithm that yields Nash Equilibrium values for the total quantity and the price. The four players pick at random one chromosome from a single population, in order to define their quantity for the current round. Then profits are calculated and the fitness value of the active chromosomes is updated, based on the profit of the player who has chosen them. The population is updated by crossover and mutation, after all chromosomes have been used. As Alkemade et al. [1] point out, the algorithm leads the total quantities and the market price to the values corresponding to the NE for these measures.

\section{The Models}

In all the above models, researchers assume symmetric cost functions (all players have identical cost functions), which implies that the Cournot games studied are symmetric. Additionally, Vriend [15], Alkemade et al. [1] and Arifovic [3] -in one of the models she investigates- use linear (and decreasing) cost functions. If a symmetric Cournot Game, has in addition, indivisibilities (discrete, but closed strategy sets), it is a pseudo-potential game [6] and the following theorem holds:

\vspace{0.2 cm}
\textbf{Theorem 1}. \textit{ ``Consider a n-player Cournot Game. We assume that the inverse demand function P is strictly decreasing and log-concave; the cost function $c_i$ of each firm is strictly increasing and left-continuous; and each firm's monopoly profit becomes negative for large enough $q$. The strategy sets $S^i$, consisting of all possible levels of output producible by firm $i$, are not required to be convex, but just closed. Under the above assumptions, the Cournot Game has a Nash Equilibrium [in pure strategies]'' }[6].

\vspace{0.2 cm}
This theorem is relevant when one investigates Cournot Game equilibrium using Genetic Algorithms, because a chromosome can have only a finite number of values and, therefore, it is the discrete version of the Cournot Game that is investigated, in principle. Of course, if one can have a dense enough discretization of the strategy space, so that the NE value of the continuous version of the Cournot Game is included in the chromosomes' accepted values, it is the case for the NE of the continuous and the discrete version under investigation to coincide.

In all three models we investigate in this paper, the assumptions of the above theorem hold, and hence there is a Nash Equilibrium in pure strategies. We investigate those models for the cases of $n=4$ and $n=20$ players.

The first model we use is the linear model used in [1]: The inverse demand is given by
\begin{equation}
P=256-Q
\end {equation}
with $Q=\sum_{i=1}^n q_i$, and the common cost function of the $n$ players is
\begin{equation}
c(q_i)=56q_i
\end{equation}
The Nash Equilibrium quantity choice of each of the 4 players is $\hat{q}=40$ [1]. In the case of 20 players we have, by solving \eqref{nash}, $\hat{q}=9.5238$.
The second model has a polynomial inverse demand function.
\begin{equation}\label{polynomialModel}
P=aQ^3-b
\end{equation}
and linear symmetric cost function
\begin{equation}\label{linearCost}
c=xq_i+y
\end{equation}
If we assume $a<0$ and $x>0$ the demand and cost functions will be decreasing and increasing, respectively, and the assumptions of theorem (1) hold. We set $a=-1$, $b=7.36\times10^7+10$, $x=y=10$, so $\hat{q}=20$ for $n=20$ and $\hat{q}=86.9401$ for $n=4$.

Finally, in the third model, we use a radical inverse demand function
\begin{equation}
P=aQ^{\frac{3}{2}} +b
\end{equation}
and the linear cost function \eqref{linearCost}. For $a=-1$, $b=8300$, $x=100$ and $y=10$ theorem (1) holds and $\hat{q}=19.3749$ for $n=20$, while $\hat{q}=82.2143$ for $n=4$.

\section{The Algorithms}
We use two multi-population (each player has its own population of chromosomes representing its alternative choices at any round) co-evolutionary genetic algorithms, Vriend's individual learning algorithm [15] and co-evolutionary programming, a similar algorithm that has been used for the game of prisoner's dilemma [10] and, unsuccessfully, for Cournot Duopoly [13]. Since those two algorithms don't, as it will be seen, lead to convergence to the NE in the models under consideration, we introduce two different versions of the algorithms, as well, which are characterized by the use of opponent choices, when the new generation of each player's chromosome population is created, and therefore can be regarded as ``socialized'' versions of the two algorithms. The difference between the ``individual'' and the ``social'' learning versions of the algorithms is that in the former case the population of each player is updated on itself (i.e. only the chromosomes of the specific player's population are taken into account when the new generation is formed), while on the latter, all chromosomes are copied into a common ``pool'',  then the usual genetic operators (crossover and mutation) are used to form the new generation of that aggregate population and finally each chromosome of the generation is copied back to its corresponding player's population. Thus we have ``social learning'', since the alternative strategic choices of a given player at a specific generation, as given by the chromosomes that comprise its population, are affected by the chromosomes  (the ideas should we say) all other players had at the previous generation.

Vriend's individual learning algorithm is presented in pseudo-code [14].
\vspace{0.2 cm}
\small
\begin{enumerate}
	\item ``A set of strategies [chromosomes representing quantities] is randomly drawn for each player.
	\item While $Period<T$
\begin{enumerate}
	\item (If $Period\ mod\ GArate = 0$): Using GA procedures \{as roulette wheel selection single, random point crossover and mutation, for generating a new set of strategies for each player [15]\}, a new set of strategies is created for each firm.
	\item Each player selects one strategy. The realized profit is calculated [and the fitness of the corresponding chromosomes, is defined, based on that profit].
\end{enumerate}
\end{enumerate}
\normalsize
\vspace{0.2 cm}
Co-evolutionary programming is quite similar, with the difference that the random match-ups between the chromosomes of the players' population at a given generation are finished when all chromosomes have participated in a game; and then the population is updated, instead of having a parameter (GArate) that defines the generations at which populations update takes place. The algorithm, described by pseudo-code, is as follows [13]:
\vspace{0.2 cm}
\small
\begin{enumerate}
	\item Initialize the strategy population of each player.
	\item Choose one strategy from the population of each player randomly, among the strategies that have not already been assigned profits. Input the strategy information to the tournament. The result of the tournament will decide profit and fitness values for these chosen strategies.
	\item Repeat step (2) until all strategies have a profit value assigned.
	\item Apply the evolutionary operators [selection, crossover, mutation] to each player's population. Keep the best strategy of the current generation alive (elitism).
	\item Repeat steps (2)-(4) until maximum number of generations has been reached.
\end{enumerate}
\normalsize
\vspace{0.2 cm}
In our implementation, we don't use elitism. The reason is that by using only selection proportional to fitness, single (random) point crossover and finally, mutation with fixed mutation rate for each chromosome bit throughout the simulation, we ensure that the algorithms can be classified as \textit{canonical economic GA's} (Riechmann 2001), and that their underlying stochastic process form an ergodic Markov Chain [12].

In order to ensure convergence to Nash Equilibrium, we introduce the two ``social'' versions of the above algorithms. Vriend's multi-population algorithm could be transformed to:
\small
\vspace{0.2 cm}
\begin{enumerate}
	\item A set of strategies [chromosomes representing quantities] is randomly drawn for each player.
	\item While $Period<T$
\begin{enumerate}
	\item (If $Period\ mod\ GArate = 0$): Use GA procedures (roulette wheel selection, single, random point crossover and mutation), to create a new generation of chromosomes, from a population consisting of the chromosomes belonging to the union of the players' populations. Copy the chromosomes of the new generation to the corresponding player's population, to form a new set of strategies for each player.
	\item Each player selects one strategy. The realized profit is calculated (and the fitness of the corresponding chromosomes, is defined, based on that profit).
\end{enumerate}
\end{enumerate}
\normalsize
\vspace{0.2 cm}
And social co-evolutionary programming is defined as:
\small
\vspace{0.2 cm}
\begin{enumerate}
	\item Initialize the strategy population of each player
	\item Choose one strategy of the population of each player randomly from among the strategies that have not already been assigned profits. Input the strategy information to the tournament. The result of the tournament will decide profit values for these chosen strategies.
	\item Repeat step (2) until all strategies are assigned a profit value.
	\item Apply the evolutionary operators (selection, crossover, mutation) at the union of players' populations. Copy the chromosomes of the new generation to the corresponding player's population to form the new set of strategies.
	\item Repeat steps (2)-(4) until maximum number of generations has been reached.
\end{enumerate}
\normalsize
\vspace{0.2 cm}
So the difference between the social and individual learning variants is that chromosomes are first copied in an aggregate population, and the new generation of chromosomes is formed from the chromosomes of this aggregate population. From an economic point of view, this means that the players take into account their opponents choices when they update their set of alternative strategies. So we have a social variant of learning, and since each player has its own population, the algorithms should be classified as ``social multi-population economic Genetic Algorithms'' [11],[12]. It is important to note that the settings of the game allow the players to observe their opponent choices after every game is played, and take them into account, consequently, when they update their strategy sets.

It is not difficult to show that the stochastic process of all the algorithms presented here form a regular Markov chain [9]. In the co-evolutionary programming algorithms (both individual and social), and since the matchings are made at random, the expected profit of the $j_{th}$ chromosome of player's $i$ population $q_{ij_i}$ is (we assume $n$ players and $K$ chromosomes in each population)
\[
E[\pi(q_{ij_i})]=
\frac{1}{(n-1)K} \sum_{j_1 =1}^K \ldots \sum_{j_{i-1} =1}^K \sum_{j_{i+1} =1}^K \ldots
\]\[ \sum_{j_n =1}^K \pi(q_{ij_i};q_{1j_1},\ldots ,q_{(i-1)(j_{i-1})},q_{(i+1)(j_{i+1})},\ldots ,q_{nj_n})
\]
The expected profit for Vriend's algorithm [14]
\[
E[\pi(q_{ij};Q_{-i})]=\bar{p}q_{ij} -C(q_{ij})
\]
with
\[
\bar{p}=\sum_{l\neq i}p(q_{ij} ,\sum_l q_{lj})f(q_{lj} |GArate)
\]
where $f(q_{ij} |GARate)$ is the frequency of each individual strategy of other firms, conditioned by the strategy selection process and GArate.

Any fitness function that is defined on the profit of the chromosomes, either proportional to profit, scaled or ordered, has a value that is solely dependent on the chromosomes of the current population. And, since the transition probabilities of the underlying stochastic process depend only on the fitness and, additionally, the state of the chain is defined by the chromosomes of the current population, the transition probabilities from one state of the GA to another, are solely dependent on the current state (see also [12]). The stochastic process of the populations is therefore, a Markov Chain. And since the final operator used in all the algorithms presented here is the mutation operator, there is a positive -and fixed- probability that any bit of the chromosomes in the population is negated. Therefore any state (set of populations) is reachable from any other state -in just one step actually- and the chain is regular.

Having a Markov chain implies that the usual performance measures -namely mean value and variance- are not adequate to perform statistical inference, since the observed values in the course of the genetic algorithm are inter-dependent. In a regular Markov chain however, one can estimate the limiting probabilies of the chain by estimating the components of the fixed frequency vector the chain converges to, by
\begin{equation}\label{freq}
\hat{\pi_i}=\frac{N_i}{N}
\end{equation}
where $N_i$ is the number of observations in which the chain is at state $i$ and $N$ is the total number of observations [4]. In the algorithms presented here, however, the number of states is extremely large. If we have $n$ players, with $k$ chromosomes consisting of $l$ bits in each player's population, the total number of possible states is $2^{knl}$, making the estimation of the limiting probabilities of all possible states, practically impossible. On the other hand, one can estimate the limiting probability of one or more given states, without needing to estimate the limiting probabilities of all the other states. A state of importance could be the state where all chromosomes of all populations represent the Nash Equilibrium quantity (which is the same for all players, since we have a symmetric game). We call this state \textit{Nash State}.

Another solution could be the introduction of \textit{lumped states} [9]. Lumped states are disjoint aggregate states consisting of more than one state, with their union being the entire space. Although the resulting stochastic process is not necessarily Markovian, the expected frequency of the lumped states can still be estimated from \eqref{freq}. The definition of the lumped states can be based on the average Hamming distance between the chromosomes in the populations and the chromosome denoting the Nash Equilibrium quantity. Denoting $q_{ij}$ the $j^{th}$ chromosome of the $i^{th}$ player's population, and $NE$ the chromosome denoting the Nash Equilibrium quantity, the Hamming distance $d(q_{ij},NE)$ between $q_{ij}$ and $NE$ would be equal to the number of bits that differ in the two chromosomes, and the average Hamming distance between the chromosomes in the populations from the Nash chromosome would be
\begin{equation}
\bar{d}=\frac{1}{nK}\sum_{i=1}^{n}\sum_{j=1}^{K} d(q_{ij},n)
\end{equation}
where $n$ is the number of players in the game and $K$ is the number of chromosomes in each player's population.We define the $i^{th}$ lumped state $S_i$ as the set of states $s_i$, in which the chromosomes' average Hamming distance from the Nash chromosome is less or equal to $i$ and greater to $i-1$

\begin{flushleft}
\textbf {Definition 1}. $S_i= \{ s_i |i-1< \bar{d}\left( q_{ij} \in s_i,n \right) \leq i \}$, for $i=1,\ldots,n$
\end{flushleft}

\vspace{0.2 cm}
The maximum value of $\bar{d}$ is equal to the maximum value of the Hamming distance between a given chromosome and the Nash chromosome. The maximum value between two chromosomes is obtained when all bits differ, and it is equal to the length of the chromosomes $L$. Therefore we have $L$ different lumped states $S_1 ,S_2 ,\ldots ,S_L$. We also define $S_0$ to be the individual Nash state (the state reached when all populations consist of the single chromosome that corresponds to the Nash Equilibrium quantity) which gives us a total of $L+1$ states. This ensures that the union of the $S_i$ is the entire populations' space, and they consist, therefore, a set of lumped states [9].
\section{Simulation Settings}
We use two variants of the three models in our simulations. One about $n=4$ players and one having $n=20$ players. We use 20-bits chromosomes for the $n=4$ players case and 8-bits chromosomes for the $n=20$ case. A usual mechanism [3],[15] is used to transform chromosome values to quantities. After an arbitrary choice for the maximum quantity, the quantity that corresponds to a given chromosome is given by: 
\begin{equation}\label{chrom2q}
q=\frac{1}{q_{max}}\sum_{k=1}^{L}q_{ijk}2^{k-1}
\end{equation}
where $L$ is the length of the chromosome and $q_{ijk}$ is the value of the $k_{th}$ bit of the given chromosome (0 or 1). According to \eqref{chrom2q} the feasible quantities belong in the interval $[0,q_{max}]$. By setting
\begin{equation}
q_{max}=3\hat{q}
\end{equation}
where $\hat{q}$ is the Nash Equilibrium quantity of the corresponding model, we ensure that the Nash Equilibrium of the continuous model is one of the feasible solutions of the discrete model, analyzed by the genetic algorithms, and that the NE of the discrete model will be therefore, the same as the one for the continuous case. And, as it can be easily proven by mathematical induction, that the chromosome corresponding to the Nash Equilibrium quantity, will always be $0101\ldots 01$, provided that chromosome length is an even number.

The $GArate$ parameter needed in the original and the ``socialized'' versions of Vriend's algorithms, is set to $GArate=50$, an efficient value suggested in the literature [15],[14]. We use single - point crossover, with the point at which chromosomes are combined [8] chosen at random. Probability of crossover is always set up to 1, i.e. all the chromosomes of a new generation are products of the crossover operation, between selected parents. The probability of mutating any single bit of a chromosome is fixed throughout any given simulation -something that ensures the homogeneity of the underlying Markov process. The values that have been used (for both cases of $n=4$ and $n=20$) are
\[
p_m= 0.1,0.075,\ldots, 0.000025, 0.00001.
\]
We used populations consisting of 
\[
pop= 20,30,40,50
\]
chromosomes. These choices were made after preliminary tests that evaluated the convergence properties of the algorithms for various population choices, and they are in accordance to the population sizes used in the literature ([15],[1], etc.).

Finally, the maximum number of generations that a given simulation runs, were
\[
T= 10^3,2*10^3,5*10^3,10^4,2*10^4,5*10^4
\]
 Note that the number of total iterations (number of games played) of Vriend's individual and social algorithms is $GArate$ times the number of generations, while in the co-evolutionary programming algorithms is number of generations times the number of chromosomes in a population, which is the number of match-ups.

We run 300 independent simulations for each set of settings for all the algorithms, so that the test statistics and the expected time to reach the Nash Equilibrium (NE state, or first game with NE played), are estimated effectively. 
\section{Presentation of Selected Results}
\begin{figure}[hp]
	\centering
		\includegraphics[width=.75\textwidth]{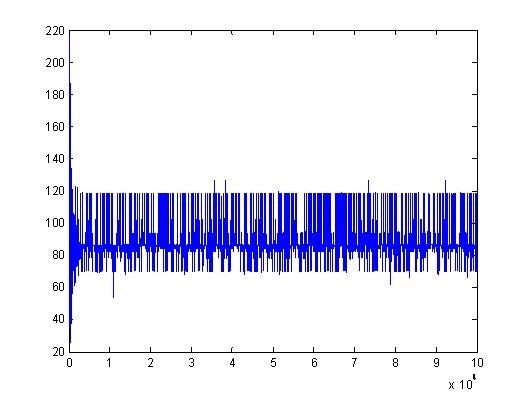}
	\caption{Mean Quantity in one execution of Vriend's individual learning algorithm in the polynomial model for $n=4$ players. $pop=50, GArate=50, p_{cr}=1, p_{mut}=0.01, T=2,000$ generations.}
	\label{fig:IndividualVriendMeanQuantity}
\end{figure}

\begin{figure}[hp]
	\centering
		\includegraphics[width=.75\textwidth]{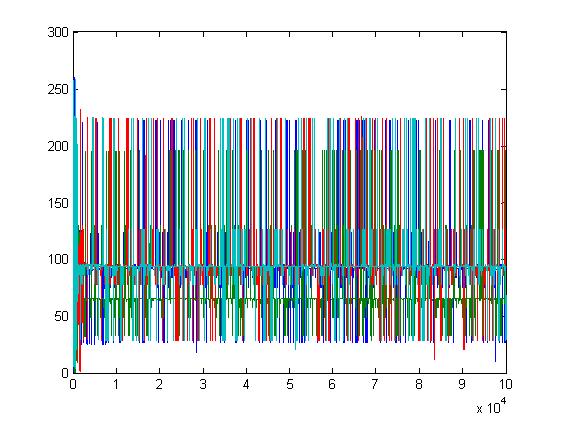}
	\caption{Players' quantities in one execution of Vriend's individual learning algorithm in the polynomial model for $n=4$ players. $pop=50, GArate=50, p_{cr}=1, p_{mut}=0.01, T=2,000$ generations.}
	\label{fig:IndividualVriendPlayersQuantities}
\end{figure}

\begin{figure}[hp]
	\centering
		\includegraphics[width=.75\textwidth]{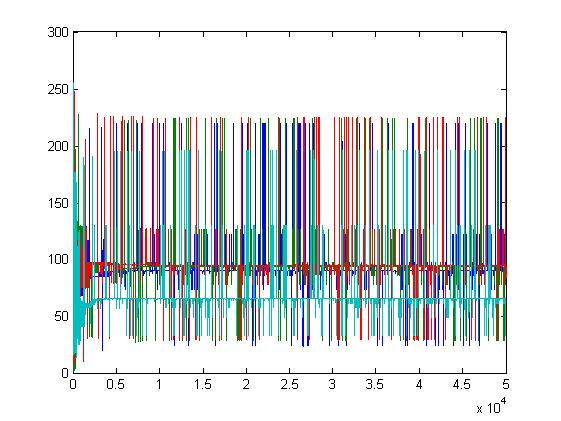}
	\caption{Players' quantities in one execution of the individual - learning version of the co-evolutionary programming algorithm in the polynomial model for $n=4$ players. $pop=50, p_{cr}=1, p_{mut}=0.01, T=2,000$ generations.}
	\label{fig:IndividualCoevolPlayersQuantities}
\end{figure}

Although the individual - learning versions of the two algorithms led the estimated expected value of the average quantity (as given in eq.\eqref{meanMeanQuantityEq})
\begin{equation}\label{meanMeanQuantityEq}
\bar{Q}=\frac{1}{nT}\sum_{t=1}^{T}\sum_{i=1}^{n}q_{it}
\end{equation}
($T=$ number of iterations, $n=$ number of players), close to the corresponding average quantity of the NE, the strategies of each one of the players converged to different quantities. That fact can be seen in figures \ref{fig:IndividualVriendMeanQuantity} to \ref{fig:IndividualCoevolPlayersQuantities}, that show the outcome of some representative runs of the two individual - learning algorithms in the polynomial model \eqref{polynomialModel}. The trajectory of the average market quantity in Vriend's algorithm 
\begin{equation}\label{meanQuantityEq}
Q=\frac{1}{n}\sum_{i=1}^{n}q_{it}
\end{equation}
(calculated in \eqref{meanQuantityEq} and shown in figure \ref{fig:IndividualVriendMeanQuantity}) is quite similar to the trajectory of the same measure in the co-evolutionary case, and a figure of the second case is omitted. The estimated average values of the two measures (eq.\eqref{meanMeanQuantityEq}) were 86.2807 and 88.5472 respectively, while the NE quantity in the polynomial model \eqref{polynomialModel} is 86.9401. The unbiased estimators for the standard deviations of the $Q$ (eq.\eqref{stdmeanQuantityEq}) were 3.9776 and 2.6838, respectively.
\begin{equation}\label{stdmeanQuantityEq}
s_{Q}=\frac{1}{T-1}\sum_{i=1}^{T}(Q_i-\bar{Q})^2
\end{equation}
The evolution of the individual players' strategies can be seen in figures \ref{fig:IndividualVriendPlayersQuantities} and \ref{fig:IndividualCoevolPlayersQuantities}. 
The estimators of the mean values of each player's quantities (calculated by eq.\eqref{meanPlayersQuantitiesEq}) 
\begin{equation}\label{meanPlayersQuantitiesEq}
\bar{q}_i=\frac{1}{T}\sum_{i=1}^{T}q_i
\end{equation}
are given on table \ref{tab:meanPlayersQuantitiesIndividualTable}, while the frequencies of the lumped states in these simulations are given on table \ref{tab:lumpedStatesndividualTable}.
\begin{table}[hp]
	\centering
\begin{tabular}{|c |c |c| }
\hline
Player&Vriend's algorithm&Co-evol. programming\\
\hline
1&91.8309&77.6752\\
2&65.3700&97.8773\\
3&93.9287&93.9287\\
4&93.9933&93.9933\\
\hline
\end{tabular}
	\caption{Mean values of players' quantities in two runs of the individual-learning algorithms in the polynomial model for $n=4$ players. $pop=50, GArate=50, p_{cr}=1, p_{mut}=0.01, T=2,000$ generations.}
	\label{tab:meanPlayersQuantitiesIndividualTable}
\end{table}

\small
\begin{table}[hp]
\begin{tabular}{|c| c c c c c c c c c c c |}
\hline
&$s_0$&$s_1$&$s_2$&$s_3$&$s_4$&$s_5$&$s_6$&$s_7$&$s_8$&$s_9$&$s_{10}$\\
VI&0&0&0&0&0&0&0&0&0&.8725&.0775\\
&$s_{11}$&$s_{12}$&$s_{13}$&$s_{14}$&$s_{15}$&$s_{16}$&$s_{17}$&$s_{18}$&$s_{19}$&$s_{20}$& \\
&.05&0&0&0&0&0&0&0&0&0&\\
\hline
&$s_0$&$s_1$&$s_2$&$s_3$&$s_4$&$s_5$&$s_6$&$s_7$&$s_8$&$s_9$&$s_{10}$\\
CP&0&0&0&0&0&0&0&0&.0025&.1178&.867\\
&$s_{11}$&$s_{12}$&$s_{13}$&$s_{14}$&$s_{15}$&$s_{16}$&$s_{17}$&$s_{18}$&$s_{19}$&$s_{20}$& \\
&.0127&0&0&0&0&0&0&0&0&0&\\
\hline
\end{tabular}	
  \caption{Lumped states frequencies in two runs of the individual-learning algorithms in the polynomial model for $n=4$ players. $pop=50, p_{cr}=1, p_{mut}=0.01, T=100,000$ generations.}
	\label{tab:lumpedStatesndividualTable}
\end{table}
\normalsize
That significant difference between the mean values of players' quantities was observed in all simulations of the individual - learning algorithms, in all models and in both $n=4$ and $n=20$, for all the parameter sets used (which were described in the previous section). We used a sample of 300 simulation runs for each parameter set and model, for hypothesis testing. The hypothesis $H_0: \bar{Q}=q_{Nash}$ was accepted for $a=.05$ in all cases. On the other hand, the hypotheses $H_0: q_i=q_{Nash}$, were rejected for all players in all models, when the probability of rejection the hypothesis, under the assumption it is correct, was $a=.05$. There was not a single Nash Equilibrium game played, in any of the simulations of the two individual - learning algorithms.

In the social - learning versions of the two algorithms, both the hypotheses $H_0: \bar{Q}=q_{Nash}$, and $H_0: q_i=q_{Nash}$ were accepted for $a=.05$, for all models and parameters sets. We used a sample of 300 different simulations for every parameter set, in those cases, as well. 

\begin{figure}[hp]
	\centering
		\includegraphics[width=.75\textwidth]{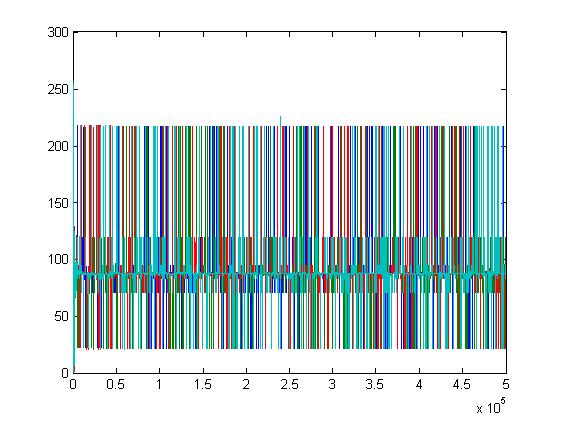}
	\caption{Players' quantities in one execution of the social - learning version of Vriend's algorithm in the polynomial model for $n=4$ players. $pop=40, GArate=50, p_{cr}=1, p_{mut}=0.00025, T=10,000$ generations.}
	\label{fig:SocialVriendPlayersQuantities}
\end{figure}

The evolution of the individual players' quantities in a given simulation of Vriend's algorithm on the polynomial model (as in fig.\ref{fig:IndividualVriendPlayersQuantities}) can be seen in fig.\ref{fig:SocialVriendPlayersQuantities}.

Notice that the all players' quantities have the same mean values (eq. \eqref{meanPlayersQuantitiesEq}). The mean values of the individual players' quantities for $pop=40, p_{cr}=1, p_{mut}=0.00025, T=10,000$ generations, are given, for one simulation of all the algorithms (social and individual versions) on table  \ref{tab:meanPlayersQuantitiesSocialTable}.

\begin{table}[hp]
\begin{tabular}{|p{1cm} |p{2.25cm} |p{2.25cm}||p{2.25cm}|p{2.25cm}| }
\hline
Player&Social&Social&Individual&Individual\\
&Vriend's alg.&Co-evol. prog.&Vriend's alg.&Co-evol. prog.\\
\hline
1&86.9991&87.0062&93.7536&97.4890\\
2&86.9905&87.0089&98.4055&74.9728\\
3&86.9994&87.0103&89.4122&82.4704\\
4&87.0046&86.9978&64.6146&90.4242\\
\hline
\end{tabular}
	\caption{Mean values of players' quantities in two runs of the social-learning algorithms in the polynomial model for $n=4$ players. $pop=40, p_{cr}=1, p_{mut}=0.00025, T=10,000$ generations.}
	\label{tab:meanPlayersQuantitiesSocialTable}
\end{table}
On the issue of establishing NE in -some- of the games played and reaching the Nash State (all chromosomes of every population equals the chromosome corresponding to the NE quantity) there are two alternative results. For one subset of the parameters set, the social - learning algorithms managed to reach the NE state and in a significant subset of the games played, all players used the NE strategy (these subsets are shown on table \ref{tab:parametersForNE}).

\begin{table}[hp]
\begin{tabular}{|p{2cm} |p{2cm} |p{1.5cm}|p{3cm}|p{1.5cm}| }
\hline
Model&Algorithm&pop&$p_{mut}$&T\\
\hline
4-Linear&Vriend&20-40&$.001-.0001$&$\geq 5000$\\
4-Linear&Co-evol&20-40&$.001-.0001$&$\geq 5000$\\
20-Linear&Vriend&20&$.00075-.0001$&$\geq 5000$\\
20-Linear&Co-evol&20&$.00075-.0001$&$\geq 5000$\\
4-poly&Vriend&20-40&$.001-.0001$&$\geq 5000$\\
4-poly&Co-evol&20-40&$.001-.0001$&$\geq 5000$\\
20-poly&Vriend&20&$.00075-.0001$&$\geq 5000$\\
20-poly&Co-evol&20&$.00075-.0001$&$\geq 5000$\\
4-radic&Vriend&20-40&$.001-.0001$&$\geq 5000$\\
4-radic&Co-evol&20-40&$.001-.0001$&$\geq 5000$\\
20-radic&Vriend&20&$.00075-.0001$&$\geq 5000$\\
20-radic&Co-evol&20&$.00075-.0001$&$\geq 5000$\\
\hline
\end{tabular}
	\caption{Parameter sets that yield NE. Holds true for both social - learning algorithms.}
	\label{tab:parametersForNE}
\end{table}
In the cases where mutation probability was too large, the ``Nash'' chromosomes were altered significantly and therefore the populations couldn't converge to the NE state (within the given iterations). On the other hand, when the mutation probability was low the number of iterations was not enough to have convergence. A larger population, requires more generations to converge to the ``NE state'' as well. The estimators of the limiting probabilities of one representative parameter set for representative cases of the first and second parameter sets are given on table \ref{tab:lumpedFreqInNE}.

\small
\begin{table}[h!]
\begin{tabular}{|p{0.5cm}|c c c c c c c c c c c |}
\hline
&$s_0$&$s_1$&$s_2$&$s_3$&$s_4$&$s_5$&$s_6$&$s_7$&$s_8$&$s_9$&$s_{10}$\\
No NE&0&0&.6448&.3286&.023&.0036&0&0&0&0&0\\
&$s_{11}$&$s_{12}$&$s_{13}$&$s_{14}$&$s_{15}$&$s_{16}$&$s_{17}$&$s_{18}$&$s_{19}$&$s_{20}$& \\
&0&0&0&0&0&0&0&0&0&0&\\
\hline
&$s_0$&$s_1$&$s_2$&$s_3$&$s_4$&$s_5$&$s_6$&$s_7$&$s_8$&$s_9$&$s_{10}$\\
NE&.261&.4332&.2543&.0515&0&0&0&0&0&0&0\\
&$s_{11}$&$s_{12}$&$s_{13}$&$s_{14}$&$s_{15}$&$s_{16}$&$s_{17}$&$s_{18}$&$s_{19}$&$s_{20}$& \\
&0&0&0&0&0&0&0&0&0&0&\\
\hline
\end{tabular}	
  \caption{Lumped states frequencies in a run of a social-learning algorithm that couldn't reach NE and another that reached it. 20 players - polynomial model, Vriend's algorithms, $pop=20$ and $T=10,000$ in both cases, $p_{mut}=.001$ in the $1^{st}$ case, $p_{mut}=.0001$ in the $2^{nd}$.}
	\label{tab:lumpedFreqInNE}
\end{table}
Apparently, the Nash state $s_0$ has greater than zero frequency in the simulations that reach it. The estimated time needed to reach Nash State (in generations), to return to it again after departing from it, and the percentage of total games played that were played on NE, are presented on table \ref{tab:markovStatisticsForNE}\footnote{Table 6: $Gen NE=$ Average number of Generations needed to reach $s_0$, starting from populations having all chromosomes equal to the opposite chromosome of the NE chromosome, in the 300 simulations. $Ret Time=$ Interarrival Times of $s_0$(average number of generations needed to return to $s_0$) in the 300 simulations. $NE Games=$ Percentage of games played that were NE in the 300 simulations.}.

\begin{table}[h!]
\begin{tabular}{|p{1.5cm} |p{1.5cm} |p{0.5cm}|p{1cm}|p{1.1cm}|p{1.2cm}|p{1cm}|p{1cm}| }
\hline
Model&Algorithm&pop&$p_{mut}$&T &Gen NE &Ret Time& NE Games\\
\hline
4-Linear&Vriend&30&$.001$&10,000&3,749.12&3.83&5.54\\
4-Linear&Co-evol&40&$.0005$&10,000&2,601.73&6.97&73.82\\
20-Linear&Vriend&20&$.0005$&20,000&2,712.45&6.83&88.98\\
20-Linear&Co-evol&20&$.0001$&20,000&2,321.32&6.53&85.64\\
4-poly&Vriend&40&$.00025$&10,000&2,483.58&3.55&83.70\\
4-poly&Co-evol&40&$.0005$&10,000&2,067.72&8.77&60.45\\
20-poly&Vriend&20&$.0005$&20,000&2,781.24&9.58&67.60\\
20-poly&Co-evol&20&$.0005$&50,000&2,297.72&,6.63&83.94\\
4-radic&Vriend&40&$.00075$&10,000&2,171.32&4.41&81.73\\
4-radic&Co-evol&40&$.0005$&10,000&2,917.92&5.83&73.69\\
20-radic&Vriend&20&$.0005$&20,000&2,136.31&7.87&75.34\\
20-radic&Co-evol&20&$.0005$&20,000&2,045.81&7.07&79.58\\
\hline
\end{tabular}
	\caption{Markov and other statistics for NE.}
	\label{tab:markovStatisticsForNE}
\end{table}

We have seen that the original individual - learning versions of the multi - population algorithms do not lead to convergence of the individual players' choices, at the Nash Equilibrium quantity. On the contrary, the ``socialized'' versions introduced here, accomplish that goal and, for a given set of parameters, establish a very frequent Nash State, making games with NE quite frequent as well, during the course of the simulations. The statistical tests employed, proved that the expected quantities chosen by players converge to the NE in the social - learning versions while that convergence cannot be achieved at the individual - learning versions of the two algorithms. Therefore it can be argued that the learning process is qualitatively better in the case of social learning. The ability of the players to take into consideration their opponents strategies, when they update theirs, and base their new choices at the totality of ideas that were used at the previous period (as in [1]), forces the strategies into consideration to converge to each other and to converge to the NE strategy as well. Of course this option would not be possible, if the profit functions of the individual players were not the same, or, to state that condition in an equivalent way, if there were no symmetry at the cost functions. If the cost functions are symmetric, a player can take note of its opponents realized strategies in the course of play, and use them as they are when he updates his ideas, since the effect of these strategies at his individual profit, will be the same. Therefore the inadequate learning process of the individually based learning can be perfected, at the symmetric case. One should note that the convergence to almost identical values displayed in the representative cases of the previous section, holds for any parameter set used in all the models presented in this paper.

The stability properties of the algorithms, are identified by the frequencies of the lumped states and the expected inter-arrival times estimated in the previous section (table \ref{tab:markovStatisticsForNE}). The inter-arrival times of the representative cases shown there are less than 10 generations. The inter-arrival times were in the same range, when the other parameter sets that yielded convergence to ``Nash state'' were used. The frequencies of the lumped states show that the 'Nash state' $s_0$ was quite frequent -for the cases it was reached, of course- and that the states defined by populations, whose chromosomes differ in less than one bits, on the average, from the Nash state itself, define the most frequent lumped state ($s_1)$. As a matter of fact the sum of these two lumped states $s_0,s_1$ was usually higher than $.90$.  As it has been already shown [4] the estimators of the limiting probabilities calculated by \eqref{freq} and presented for given simulation runs, on tables \ref{tab:lumpedStatesndividualTable} and \ref{tab:lumpedFreqInNE}, are unbiased and efficient estimators for the expected frequencies of the algorithm's performance ad infinitum. The high expected frequencies of the lumped states that are ``near'' the NE and the low inter-arrival time to the NE state itself, ensure the stability of the algorithms.

Using these two algorithms as heuristics to discover unknown NE, requires a way to distinguish the potential Nash Equilibrium chromosomes. When VS\footnote{Social - learning version of Vriend's algorithm} or CS\footnote{Social - learning version of co - evolutionary programming}  converge -in the sense mentioned above- to the ``Nash state'', most chromosomes in the populations of several of the generations at the end of the simulation, should be identical or almost identical (differing at a small number of bits) to the Nash Equilibrium chromosome. Using this qualitatively rule, one should be able to find some potential chromosomes to check for Nash Equilibrium. A more concise way, would be to record the games that all players used the same quantities. Since symmetric profits functions imply symmetric NE, apparently, one can confine his attention on these games, of all the games played. In order to check if any of these quantities is the NE quantity, one could assume that all but one players use that quantity and then solve (either analytically, numerically or by a heuristic, depending on the complexity of the model investigated) the single - variable maximization problem for the player's profit, given that the other players choose the quantity under consideration. If the solution of the problem is the same quantity, then that quantity should be the Nash Equilibrium.
\section {Conclusions}
We have seen that the social-learning multi-population algorithms introduced here lead to convergence of the individual quantities to the Nash Equilibrium quantity on several Cournot models. That convergence was achieved for given parameter sets (mutation probability, number of generations, etc.) and was true in  a ``Ljapunov'' sense, i.e. the strategies chosen fluctuated inside a region around the NE, while the expected values were equal (as proven by a series of statistical tests) to the desired value. This property, which does not hold for the individual - learning variants of the two algorithms, allows one to construct heuristic algorithms to discover an unknown Nash Equilibrium in symmetric games, provided the parameters used are suitable and that the NE belongs in the feasible set of the chromosomes' values. 
Finally, the stability properties of the social-learning versions of the algorithms allow one to use them as modeling tools in a multi - agent learning environment, that leads to effective learning of the Nash Strategy.
Paths for future research could be simulating these algorithms for different bit-lengths of the chromosomes in the populations since, apparently, the use of more bits for chromosome encoding implies more feasible values for the chromosomes and, therefore, makes the inclusion of unknown NE in these sets, more probable. Another idea would be to use different models, especially models that do not have single NE. Finally one could try to apply the algorithms introduced here in different game theoretic problems.

\vspace{0.3 cm}
\begin{flushleft}
\textbf{Aknowledgements}
\end{flushleft}

Funding by the EU Commission through COMISEF MRTN-CT-2006-034270 is gratefully acknowledged. Mattheos Protopapas would also like to thank Professors Peter Winker, Manfred Gilli, Dietmar Maringer and Thomas Wagner for their extremely helpful courses.

\vspace{0.3 cm}
\begin{flushleft}
\textbf{References}
\end{flushleft}
  \begin{itemize}
  \item[{[1]}] Alkemade F, La Poutre H, Amman H (2007) On Social Learning and Robust Evolutionary Algorithm Design in the Cournot Oligopoly Game. Comput Intell 23: 162--175.
  \item[{[2]}] Alos-Ferrer C, Ania A (2005) The Evolutionary Stability of Perfectly Competitive Behavior. Econ Theor 26: 497--516.
  \item[{[3]}] Arifovic J (1994) Genetic Algorithm Learning and the Cobweb Model. J Econ Dynam Contr 18: 3--28.
  \item[{[4]}] Basawa IV, Rao P (1980) Statistical Inference for Stochastic Processes. Academic Press, London.
  \item[{[5]}] Dawid H, Kopel M (1998) On Economic Applications of the Genetic Algorithm: A Model of the Cobweb Type. J Evol Econ 8: 297--315.
  \item [{[6]}] Dubey P, Haimanko O, Zapechelnyuk A (2006) Strategic Complements and Subtitutes and Potential Games. Game Econ Behav 54: 77--94.
  \item [{[7]}] Franke R (1998) Coevolution and Stable Adjustments in the Cobweb Model. J Evol Econ 8: 383--406.
  \item [{[8]}] Goldberg DE (1989) Genetic Algorithms in Search, Optimization and Machine Learning. Addison - Wesley, Reading MA.
  \item [{[9]}] Kemeny J, Snell J (1960) Finite Markov Chains. D.Van Nostrand Company Inc., Princeton MA.
  \item[{[10]}]  Price TC (1997) Using Co-Evolutionary Programming to Simulate Strategic Behavior in Markets. J Evol Econ 7: 219--254.
  \item [{[11]}] Riechmann T (1999) Learning and Behavioral Stability. J Evol Econ 9: 225--242.
  \item[{[12]}] Riechmann T (2001) Genetic Algorithm Learning and Evolutionary Games. J Econ Dynam Contr 25: 1019--1037.
  \item[{[13]}] Son YS, Baldick R (2004) Hybrid Coevolutionary Programming for Nash Equilibrium Search in Games with Local Optima. IEEE Trans Evol Comput 8: 305--315.
  \item [{[14]}] Vallee T, Yildizoglou M (2007) Convergence in Finite Cournot Oligopoly with Social and Individual Learning. Working Papers of GRETha, 2007-07. Available by GRETha ( http://www.gretha.fr ) Accessed 10 November 2007.
  \item [{[15]}] Vriend N (2000) An Illustration of the Essential Difference between Individual and Social Learning, and its Consequences for Computational Analyses. J Econ Dynam Contr 24: 1--19.
  \end{itemize}
\end{document}